# Transfer Learning with Deep Convolutional Neural Network (CNN) for Pneumonia Detection using Chest X-ray

Tawsifur Rahman[1], Muhammad E. H. Chowdhury[2], Amith Khandakar[2], Khandaker R. Islam[3], Khandaker F. Islam[2], Zaid B. Mahbub[4], Muhammad A. Kadir[1], Saad Kashem[5]

[1]Department of Biomedical Physics & Technology, University of Dhaka, Dhaka-1000, Bangladesh

[2]Department of Electrical Engineering, Qatar University, Doha-2713, Qatar

[3]Department of Orthodentics, Bangabandhu Sheikh Mujib Medical University, Dhaka-1000, Bangladesh

[4]Department of Mathematics and Physics, North South University, Dhaka-1229, Bangladesh

[5]Faculty of Robotics and Advanced Computing, Qatar Armed Forces-Academic Bridge Program, Qatar Foundation, Doha-24404, Qatar

**\*Correspondence:** Dr. Muhammad E. H. Chowdhury; Tel.: +974-31010775

**Abstract:** Pneumonia is a life threatening disease, which occurs in the lungs caused by either bacterial or viral infection. It can be life endangering if not acted upon in the right time and thus early diagnosis of pneumonia is vital. The aim of this paper is to automatically detect bacterial and viral pneumonia using the digital x-ray images. It provides a detailed report on advances made in making accurate detection of pneumonia and then presents the methodology adopted by the authors. Four different pre-trained deep Convolutional Neural Network (CNN): AlexNet, ResNet18, DenseNet201 and SqueezeNet were used for transfer learning. 5247 chest x-ray images consisting of bacterial, viral and normal chest x-rays were used and through preprocessing techniques, the modified images were trained for the transfer learning based classification task. In this work, the authors have reported three schemes of classifications: normal vs pneumonia, bacterial vs viral pneumonia and normal, bacterial and viral pneumonia. The classification accuracy of normal and pneumonia images, bacterial and viral pneumonia images, and normal, bacterial and viral pneumonia were 98%, 95% and 93.3% respectively. This is the highest accuracy in any scheme than the accuracies reported in the literature. Therefore, the proposed study can be useful in faster diagnosing pneumonia by the radiologist and can help in the fast airports screening of the pneumonia patients.

## 1. Introduction

Pneumonia is considered the greatest cause of children in all over the world. Approximately 1.4 million children die of pneumonia every year, which is 18% of the total children died at less than five years old [1]. Globally overall two billion people are suffering from pneumonia every year [1].

Pneumonia is a lung infection, which can be caused by either bacteria or viruses. Luckily, this bacterial or viral infectious disease can be well treated by antibiotics and antivirals drugs. Nevertheless, faster diagnosis of viral or bacterial pneumonia and consequent application of correct medication can help significantly to prevent deterioration of the patient condition which eventually leads to death [2]. Chest X-rays are currently the best method for diagnosing pneumonia [3]. X-ray images of pneumonia is not very clear and often misclassified to other diseases or other benign abnormalities. Moreover, the bacterial or viral pneumonia images are sometimes miss-classified by the experts, which leads to wrong medication to the patients and thereby worsening the condition of the patients [4-6]. There are considerable subjective inconsistencies in the decisions of radiologists were reported in diagnosing pneumonia. There is also a lack of trained radiologist in low resource countries (LRC) especially in rural areas. Therefore, there is a pressing need for computer aided diagnosis (CAD) systems, which can help the radiologists in detecting different types of pneumonia from the chest X-ray images immediately after the acquisition.

Currently many biomedical complications (e.g., brain tumor detection, breast cancer detection, etc.) are using Artificial Intelligence (AI) based solutions [7-10]. Among the deep learning techniques, convolutional neural networks (CNNs) have shown great promise in image classification and therefore widely adopted by the research community [11] Deep Learning Machine learning techniques on chest X-Rays are getting popularity as they can be easily used with low-cost imaging techniques and there is an abundance of data available for training different machine-learning models. Several research groups [1, 12-25] have reported the use of deep machine learning algorithms in the detection of pneumonia, however only an article [16] has reported the classification of bacterial and viral pneumonia.

There are many works where the authors tried varying the parameters of deep layered CNN for pneumonia detection. The pattern of diffuse opacification in the lung radiograph due to pneumonia can be alveolar or interstitial. Patients with alveolar infiltration on the chest radiograph, especially those with lobar infiltrates, have laboratory evidence of a bacterial infection [26]. Again, interstitial infiltrations on radiograph may be linked with viral pneumonia [27]. These might be the distinctive features in the machine learning algorithms in differentiating viral and bacterial pneumonia. Some researchers have promising results such as Liang *et al.* [18], Vikash *et al.* [28], Krishnan *et al.* [29] and Xianghong *et al.* [30]. Some groups have combined CNN with rib suppression and lung filed segmentation for other classification tasks while some highlighted a visualization technique in CNN to find the region of interest (ROI) that can be used to identify pneumonia and distinguish between bacterial and viral types in pediatric Chest X-rays. Concept of transfer learning in deep learning framework was used by Vikash *et al.* [28] for the detection of pneumonia using pre-trained ImageNet models[31] and their ensembles. A customized VGG16 model was used in Xianghong et al. [30], which consists of two parts, lung regions identification with a fully convolutional network (FCN) model and pneumonia category classification using a deep convolutional neural network (DCNN). A dataset containing 32,717 patients' X-rays were used in a deep learning technique by Wang *et al.* [12] producing promising results. Ronneburger *et al.* [13] further went on using data augmentation along with CNN to get better results by training on small set of images. Rajpurkar *et al.* [32] reported a 121-layer CNN on chest X-rays to detect 14 different pathologies, including pneumonia using an ensemble of different networks. Jung *et al.* [33] used a 3D Deep CNN with shortcuts and dense connections which help in solving the gradient vanishing problem. Jaiswal *et al.* [17] and Siraz *et* *al.* [34] used a deep neural network called Mask-RCNN for segmentation of pulmonary images. This was used along with image augmentation for pneumonia identification, which was ensembled with RetinaNet to localize pneumonia. A pretrained DenseNet-121 and feature extraction techniques were used in the accurate identification of 14 thoracic diseases in [16]. Sundaram et al. [35] used AlexNet and GoogLeNet [36] with data augmentation to obtain an Area Under the Curve (AUC) of 0.94–0.95. The same authors had better results of AUC of 0.99 using modified two-network ensemble architecture.

The highest accuracy reported in the above-mentioned literatures in classifying the normal vs pneumonia patients and bacterial vs viral pneumonia using X-ray images using deep learning algorithms was 96.84% and 93.6% respectively. Therefore, there is significant room for improving the result either by using different deep learning algorithms or modifying the existing outperforming algorithms or combining several outperforming algorithms as an ensemble model to produce a better classification accuracy particularly in classifying viral and bacterial pneumonia. The proposed study is reporting a transfer learning approach using four different pre-trained network architectures (AlexNet, ResNet18, DenseNet201, and SqueezeNet) and analyzed their performances. The key contribution of this work is to provide a CNN based transfer-learning approach using different pre-trained models to detect pneumonia and classify bacterial and viral pneumonia with higher accuracy compared to the recent works. In addition, the paper also provides the methodological details of the work, which can be utilized by any research group to take the benefit of this work. Moreover, radiographic findings are poor indicators for the diagnosis of the cause of pneumonia until now. So the motivation of the present study was to utilize the power of machine learning, firstly to diagnose pneumonia by analyzing radiograph and secondly to differentiate viral and bacterial pneumonia with better accuracy.

The rest of the paper is divided in the following section: Section 2 summarizes different pre-trained networks used for this study and Section 3 describes the methodology used in the study, where the details of the dataset and the pre-processing steps to prepare the data for training and testing are provided. Section 4 provides the results of the classification algorithms, which is compared with some other recent studies while results are discussed in Section 5 and finally, the conclusion is presented in Section 6.

## 2. Background of Deep Machine Learning Algorithms

### *2.1 Convolutional Neural Networks (CNNs)*

As discussed earlier, CNNs have been popular due to their improved performance in image classification. The convolutional layers in the network along with filters help in extracting the spatial and temporal features in an image. The layers have a weight-sharing technique, which helps in reducing computation efforts [37] [38].

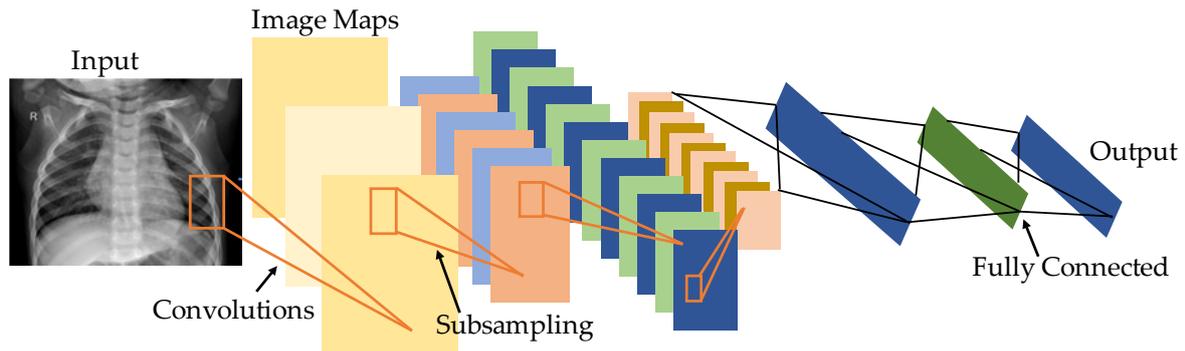

**Figure 1:** CNN Architecture.

Architecture wise, CNNs are simply feedforward artificial neural networks (ANNs) with two constraints: neurons in the same filter are only connected to local patches of the image to preserve spatial structure and their weights are shared to reduce the total number of the model's parameters. A CNN consists of three building blocks: i) Convolution layer to learn features, ii) Max-Pooling (subsampling) layer is to down sample the image and reduce the dimensionality and thereby reduction in computational efforts, and iii) Fully connected layer to equip the network with classification capabilities [39]. The architectural overview of CNN is illustrated in Figure 1.

## 2.2 *Deep Transfer learning*

CNNs typically outperforms in a larger dataset than a smaller one. Transfer learning can be useful in those applications of CNN where the dataset is not large. The concept of transfer learning is shown in the figure 2, where the trained model from large dataset such as ImageNet [40] can be used for application with comparatively smaller dataset.

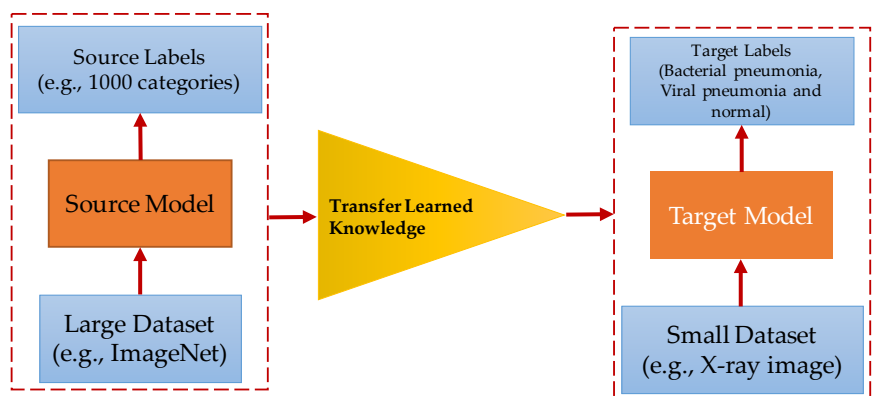

**Figure 2**: Concept of Transfer Learning [41].

Recently transfer learning has been successfully used in various field applications such as manufacturing, medical and baggage screening [42-44]. This removes the requirement of having large dataset and also reduces the long training period as is required by the deep learning algorithm when developed from scratch [45, 46].

## 2.3 *Pre-trained Convolutional Neural Networks*

In this study, four well-known pre-trained deep learning CNNs: AlexNet [11], ResNet [47], DenseNet [47] & SqueezeNet [48] were used for pneumonia detection. Brief introduction of these pre-trained networks is mentioned below:

**AlexNet**

AlexNet can classify more than 1000 different classes using deep layers consisting 650k neurons and 60 million parameters. The network is made up of 5 convolutional layers (CLs) with three pooling layers, 2 fully connected layer (FLCs) and a Softmax layer [11]. The dimension of input image for the AlexNet has to be 227×227×3 and the first CL converts input image with 96 kernels sized at 11×11×3 having a stride of 4 pixels, which is the input to second layer [49] and the remaining details are summarized in Figure 3.

**ResNet18**

ResNet which is originated from Residual Network (Figure 4), was originally developed to two problems such as vanishing gradient and degradation problem [47]. Residual learning tries to solve both these problems. ResNet has three different variants: ResNet18, ResNet50 and ResNet101 based on the number of layers in the residual network. ResNet was successfully used in biomedical image classification [48] for transfer learning. In this paper, we have used ResNet18 for the pneumonia detection. Typically, deep neural network layers learn low or high level features during training while ResNet learns residuals instead of features.

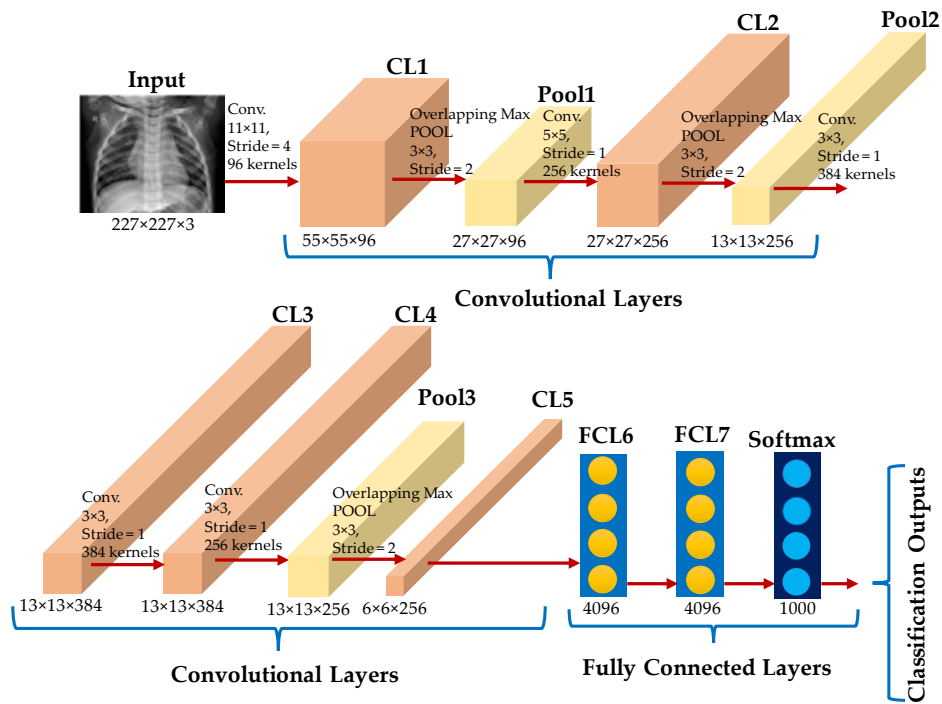

**Figure 3**: AlexNet Structure [11, 41].

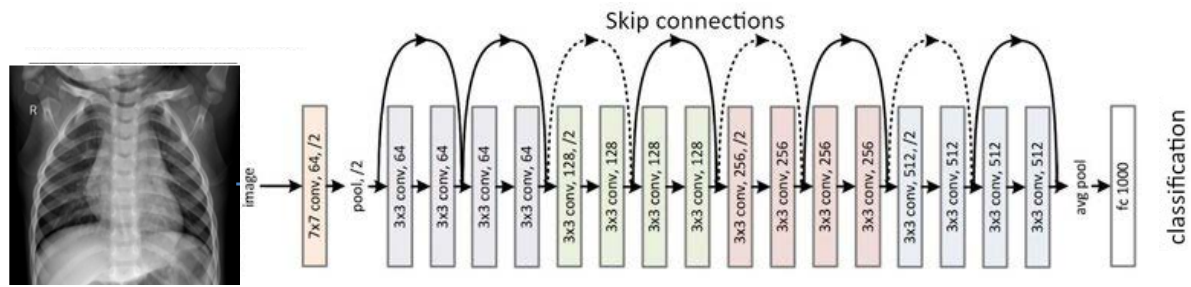

**Figure 4:** Structure of ResNet18 [48].

**DenseNet201**

DenseNet, which is a short form of Dense Convolutional Network, needs less number of parameters than a conventional CNN as it does not learn redundant feature maps. The layers in DenseNet are very narrow, i.e., 12 filters, which add a lesser set of new feature-maps. DenseNet has four different variants: DenseNet121, DenseNet169, DenseNet201 and DenseNet264. In this paper, we have used DenseNet201 for the pneumonia detection [50]. Each layer in DenseNet (as shown in Figure 5) has direct access to the original input image and gradients from the loss function. Therefore, the computational cost significantly reduced, which makes DenseNet a better choice for image classification.

**SqueezeNet**

SqueezeNet [51] is another CNN, which was trained using ImageNet database [51]. The SqueezeNet was trained with more than 1 million images and it has 50 times fewer parameters than AlexNet. The foundation of this network is a fire module, which consists of Squeeze Layer and Expand layer. The Squeeze layer has only 1 × 1 filters, which is feeding to an Expand layer than has a mixture of 1 × 1 and 3 × 3 convolution filters [51] (Figure 6). We have used SqueezeNet pre-trained model to detect pneumonia and classify bacterial and viral pneumonia in this research.

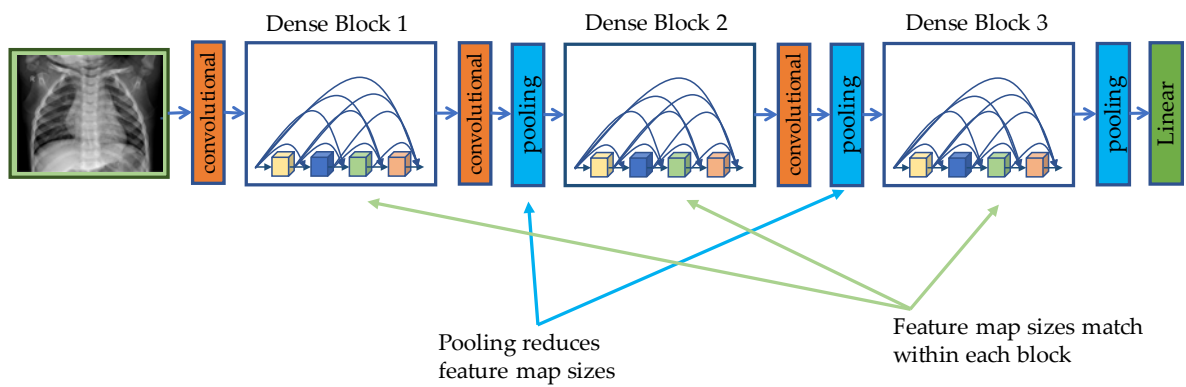

**Figure 5:** DenseNet201 architecture [52]

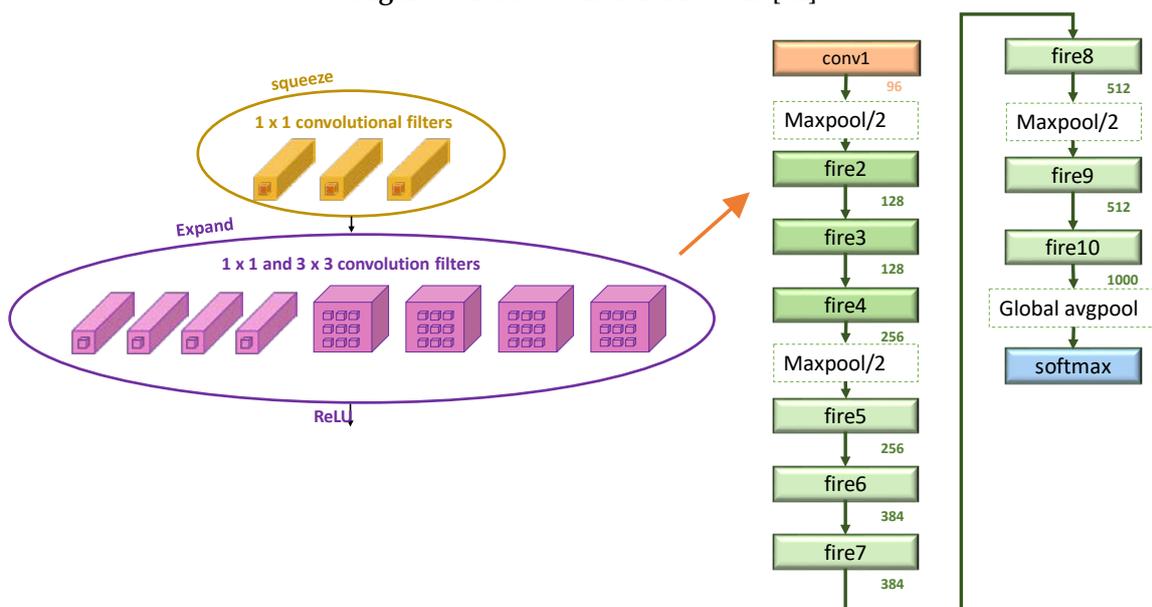

**Figure 6:** SqueezeNet architecture [51].

## 3. Methodology

### *3.1 Dataset*

In this work, kaggle chest X-ray pneumonia database was used, which is comprised of 5247 chest X-ray images with resolution varying from 400p to 2000p [53]. Out of 5247 chest X-ray images, 3906 images are from different subjects affected by pneumonia (2561 images for bacterial pneumonia and 1345 images for viral pneumonia) and 1341 images are from normal subjects (Table 1). Mixed viral and bacterial infection occurs in some cases of pneumonia. However, the dataset used in this study does not include any case of viral and bacterial co-infection. This dataset was segmented into training and test set.

**Table 1:** Complete Dataset details

| Type | Number of X-ray images |
|---|---|
| Normal | 1341 |
| Bacterial Pneumonia | 2561 |
| Viral Pneumonia | 1345 |
| **Total** | **5247** |

Table 2 shows number of train (using augmentation) and test images for different evaluation experiments. Four different algorithms were trained using the training dataset and then evaluated on test dataset. Figure 7 shows two sample for normal, bacterial and viral pneumonia chest X-ray images.

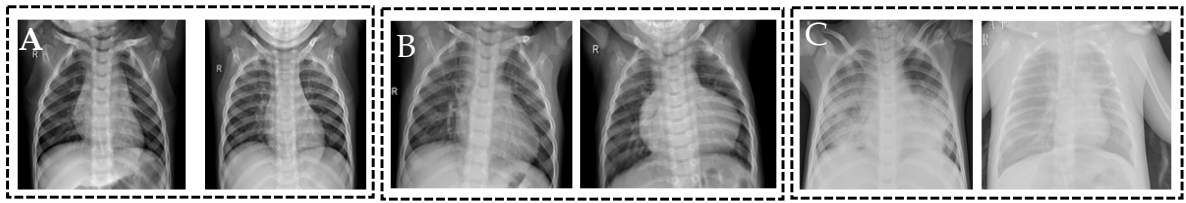

**Figure 7:** Data samples from the dataset, (A) shows normal cases, (B) shows bacterial pneumonia cases and (C) shows viral pneumonia cases.

**Table 2:** Details of Training and Test set.

| Types | | Training Set (using augmentation) | Test Set |
|---|---|---|---|
| Normal and Pneumonia | Normal | 4500 | 205 |
| | Pneumonia | 4500 | 214 |
| Normal, Bacterial, and Viral Pneumonia | Normal | 4500 | 199 |
| | Bacterial Pneumonia | 4500 | 197 |
| | Viral Pneumonia | 4500 | 201 |
| Bacterial and Viral Pneumonia | Bacterial Pneumonia | 4500 | 197 |
| | Viral Pneumonia | 4500 | 201 |

In this study, MATLAB (2019a) was utilized to train, evaluate and test different algorithm. Figure 8 illustrates the overview of the methodology of this study. Image sets undergo some pre-processing steps and data augmentation and then training using pre-trained algorithms: AlexNet, ResNet18, DenseNet201, and SqueezeNet and tested all the algorithms on the test dataset. The training of the different models was carried out in a computer with Intel© i7-core @3.6GHz

processor and 16GB RAM, 2 GB graphics card with graphics processing unit (GPU) on 64-bit Windows 10 operating system. Different parameters used for training the CNN models were shown in the Table 3. ResNet and DenseNet have different variants however, ResNet18 and DenseNet201 were readily available for Matlab 2019a and therefore used in this study. Moreover, ResNet18 outperforms other ResNet models in terms of radiographic classification.

**Table 3:** Training parameters of different pre-trained CNN models.

| Software | Pre-trained CNN models | Image size | Optimization | Momentum | Mini-batch | Learning rate |
|---|---|---|---|---|---|---|
| MATLAB (2029a) | AlexNet | 227×227 | Gradient descent | 0.9 | 16 | 0.0003 |
| | ResNet18 | 224×224 | | | | |
| | DenseNet201 | 224×224 | | | | |
| | SqueezeNet | 227×227 | | | | |

*3.2 Preprocessing*

One of the important steps in the data preprocessing was to resize the X-Ray images as the image input for different algorithms were different. For AlexNet and SqueezeNet, the images were resized to 227×227 pixels whereas for ResNet18 and DenseNet201, the images were resized to and 224×224 pixels. All images were normalized according to the pre-trained model standards.

*Data augmentation*

As discussed earlier, CNNs work better with large dataset. However, the size of the working database is not very large. There is a common trend in training deep learning algorithms to make the comparatively smaller dataset to a large one using data augmentation techniques. It is reported that the data augmentation can improve the classification accuracy of the deep learning algorithms. The performance of the deep learning models can be improved by augmenting the existing data rather than collecting new data. Some of the deep learning frameworks have data augmentation facility built-in the algorithms, however, in this study, authors have utilized three augmentation strategies to generate new training sets (Rotation, Scaling, and Translation) shown in Figure 9.

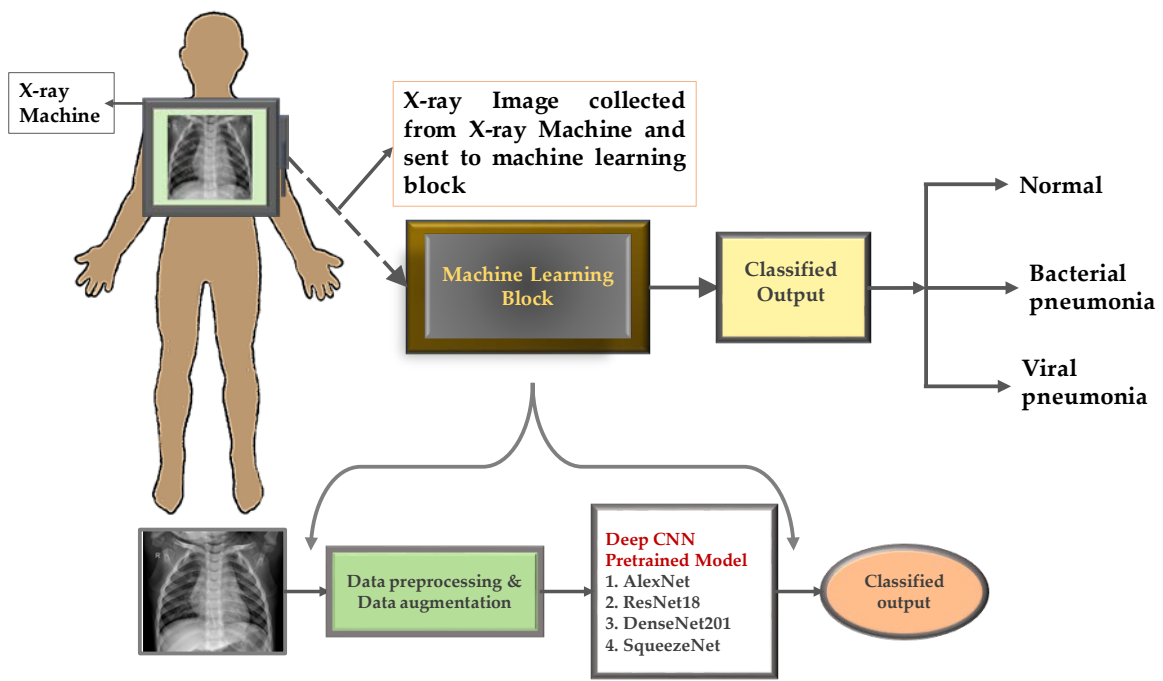

**Figure 8:** Overview of the methodology.

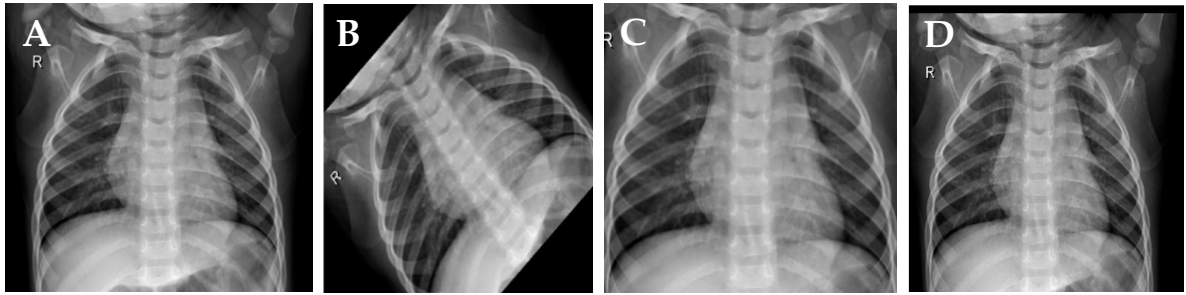

**Figure 9**: Original Chest-X-ray image (A), Chest-X-ray image after rotation (B), Chest-X-ray image after scaling (C), Chest-X-ray image after translation (D).

The rotation operation used for image augmentation is normally done by rotating the image in the clockwise direction by an angle between 0 to 360 degrees, which rotates the pixel of the image frame and fill the area of the image where there was no image pixel. In this work, the rotation of 315 degrees (counter clockwise 45 degrees) was used. The scaling operation is the magnification or reduction of frame size of the image, which is another augmentation technique used. 10% of image magnification as done as shown in Figure 9(C). Image translation can be done by translating the image in either horizontal (width shift) or vertical direction (height shift) or in both directions. Original image was horizontally translated by 10% and vertically translated by 010%.

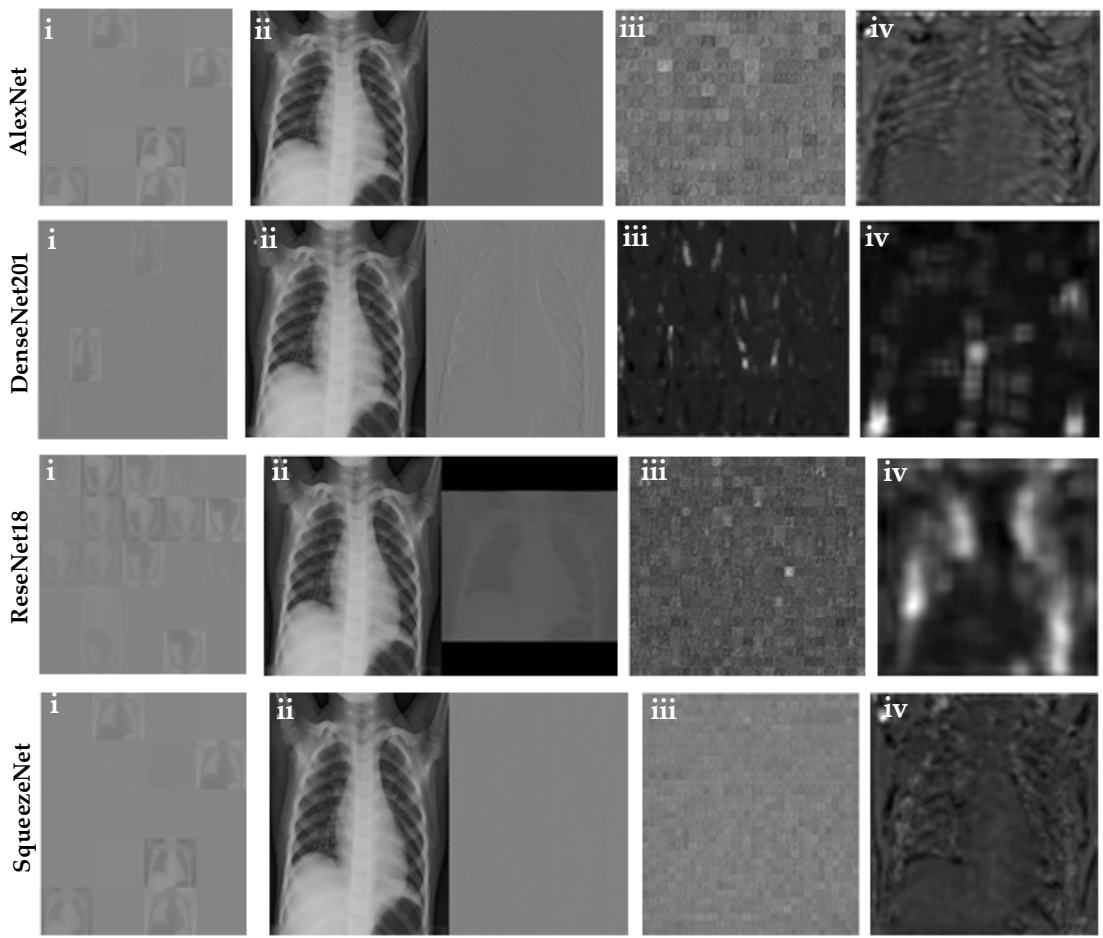

**Figure 10:** Activation map for different network models of (i) First convolutional layer, (ii) Strongest activation channel, (iii) Deep layer images set, and (iv) Deep convolutional layer in specific image.

### 3.3 *Visualization of the Activation Layer*

We investigated the features of the image by observing which areas in the convolutional layers activated on an image by comparing with the matching regions in the original images. Each layer of a CNN consists of many 2-D arrays called *channels*. The input image was applied to different networks and the output activations of the first convolution layer was examined.

The activations for different network models is shown in Figure 10. The activation map can take different range values and was therefore normalized between 0 and 1. The strongest activation channels were observed and compared with the original image. It was noticed that this channel activates on edges. It activates positively on light left/dark right edges, and negatively on dark left/light right edges.

Most convolutional neural networks learn to detect features like color and edges in their first convolutional layer. In deeper convolutional layers, the network learns to detect features that are more complicated. Later layers build up their features by combining features of earlier layers. Figure 10 shows the activation map in early convolutional layers, deep convolutional layer and strongest activation channel for each of the models.

### 3.4 *Different Experiments*

Three different forms of performance evaluations and comparisons were carried out in this study: two classes (normal and pneumonia), three classes (normal, bacterial pneumonia and viral pneumonia), and two classes (bacterial pneumonia and viral pneumonia) classification using four different deep learning algorithms through transfer learning.

The experiment carried out in this study consists of three steps. In the first step, the dataset divided into normal and pneumonia. Second step, the dataset divided into normal, bacterial and viral pneumonia. Last step, the dataset divided into bacterial and viral pneumonia. An end-to-end training approach was adopted to classify normal, bacterial and viral pneumonia images.

### 3.5 *Performance Matrix for Classification*

Four CNNs were trained and evaluated using five fold cross-validation in this work. The performance of different networks for testing dataset is evaluated after the completion of training phase and was compared using six performances metrics such as- accuracy, sensitivity or recall, Specificity, Precision (PPV), Area under curve (AUC), F1 score. Table 3 shows six performance metrics for different deep CNNs:

$$Accuracy = \frac{(TP+TN)}{(TP+FN)+(FP+TN)} \qquad (1)$$

$$Sensitivity = \frac{(TP)}{(TP+FN)} \qquad (2)$$

$$Specificity = \frac{(TN)}{(FP+TN)} \qquad (3)$$

$$Precision = \frac{(TP)}{(TN+FP)} \qquad (4)$$

$$F1\ Score = \frac{(2*TP)}{(2*TP+FN+FP)} \qquad (5)$$

In the above equations, while classifying normal and pneumonia patients, true positive (TP), true negative (TN), false positive (FP) and false negative (FN) were used to denote number of pneumonia images identified as pneumonia, number of normal images identified as normal, number normal images incorrectly identified as pneumonia images and number of pneumonia images incorrectly identified as normal, respectively. On the other hand, while classifying viral and bacterial pneumonia, TP, TN, FP, and FN were used to denote number of viral pneumonia images identified as viral pneumonia, number of bacterial pneumonia images identified as bacterial pneumonia, number bacterial pneumonia images incorrectly identified as viral pneumonia images and number of viral pneumonia images incorrectly identified as bacterial pneumonia, respectively.

## 4. Results and Discussions

The comparative performance of training and testing accuracy for different CNNs for classification schemes were shown in Figure 11. It can be noted that for three classification schemes DenseNet201 is producing the highest accuracy for both training and testing. For normal and pneumonia classification, the test accuracy was 98%, while for normal, bacterial and viral pneumonia classification, it was 93.3%, and for bacterial and viral pneumonia classification, it was found to be 95%. Figure 12 shows the area under the curve (AUC) /receiver-operating characteristics (ROC) curve (also known as AUROC (area under the receiver operating characteristics)) for different classification schemes, which is one of the most important evaluation metrics for checking any classification model's performance. This is also evident from Figure 12 that DenseNet201 outperforms the other algorithms.

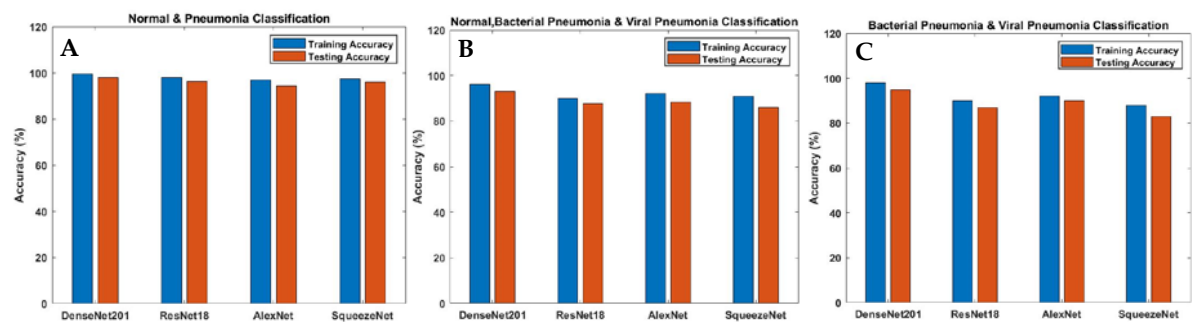

**Figure 11:** Comparison of training and testing accuracy for Normal and Pneumonia (A), Normal, bacterial and viral Pneumonia (B), Bacterial and viral Pneumonia (C) classification for different models.

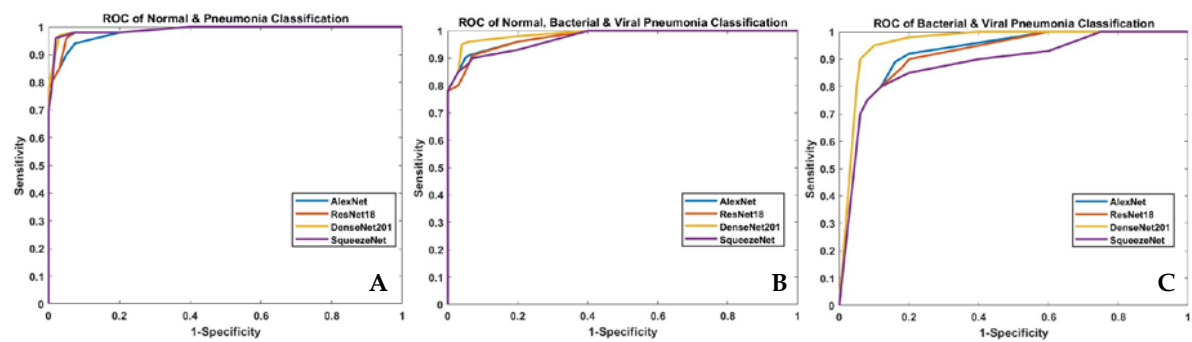

**Figure 12:** Comparison of the ROC curve for Normal and Pneumonia (A), Normal, bacterial and viral Pneumonia (B), and Bacterial and viral Pneumonia classification using CNN based models.

Figure 13 shows the confusion matrix for outperforming Densenet201 pre-trained model in for three different classifications.

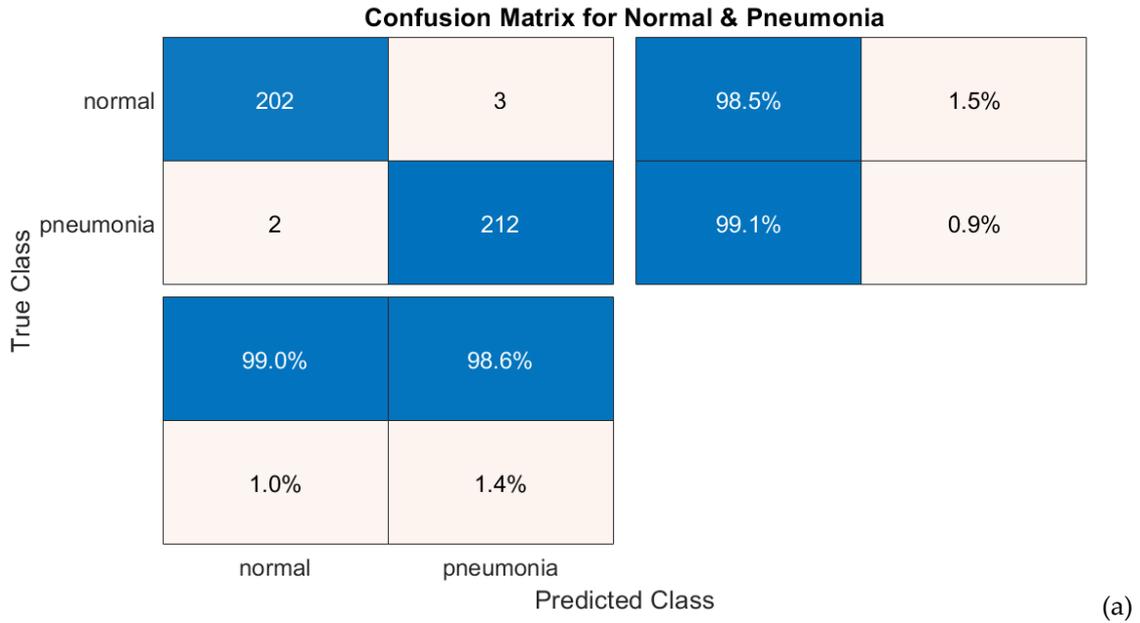

(a)

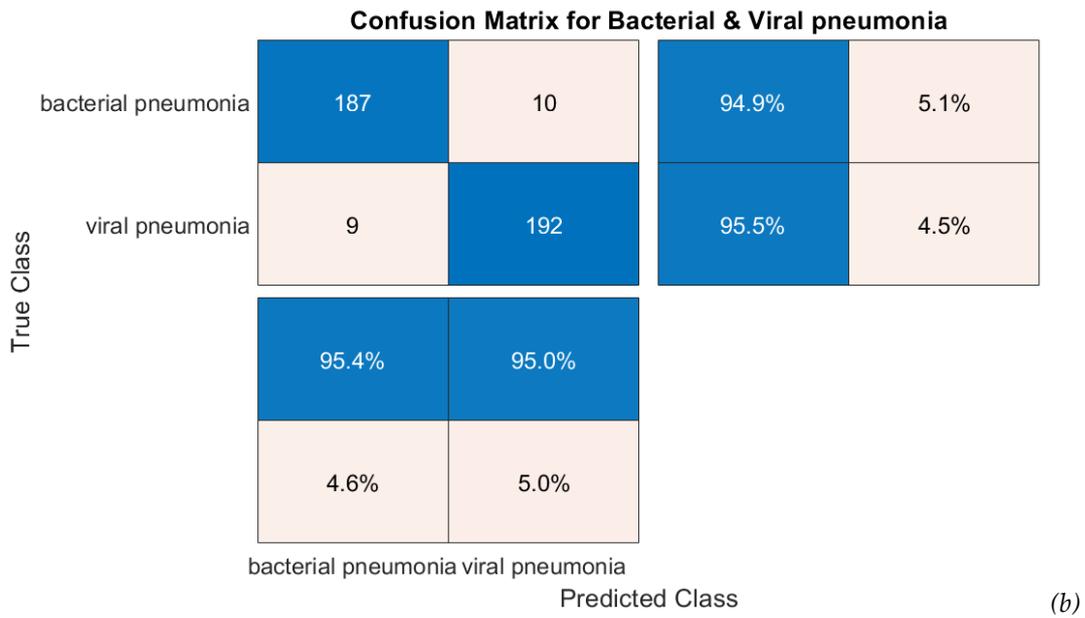

(b)

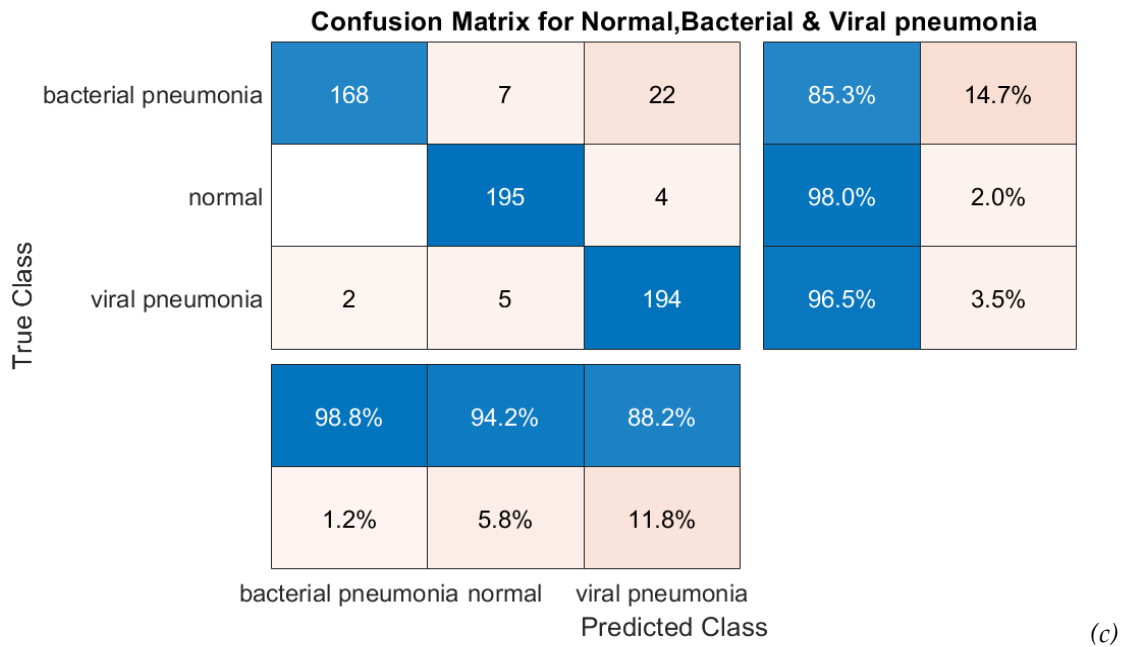

*(c)*

**Figure 13:** Confusion matrix for (a) Normal and Pneumonia, (b) Bacterial and viral, (c) Normal, bacterial and viral Pneumonia classification using DenseNet201.

Table 4 summarizes the performance matrix for different CNN algorithms tested for the three different classification schemes. DenseNet201 outperforms other models in three different classification schemes in terms of different performance indices.

**Table 4:** Different performance metrics for different deep learning networks.

| Task | Models | Accuracy | Sensitivity (Recall) | Specificity | Precision (PPV) | Area under curve (AUC) | F1 Scores |
|---|---|---|---|---|---|---|---|
| Normal and Pneumonia | AlexNet | 0.945 | 0.953 | 0.926 | 0.931 | 0.942 | 0.943 |
| | ResNet18 | 0.964 | 0.97 | 0.95 | 0.954 | 0.963 | 0.965 |
| | DenseNet201 | **0.98** | 0.99 | 0.97 | 0.97 | 0.98 | 0.981 |
| | SqueezeNet | 0.961 | 0.94 | 0.98 | 0.985 | 0.96 | 0.961 |
| Normal, Bacterial Pneumonia and Viral Pneumonia | AlexNet | 0.884 | 0.883 | 0.941 | 0.886 | 0.911 | 0.885 |
| | ResNet18 | 0.877 | 0.88 | 0.94 | 0.875 | 0.91 | 0.909 |
| | DenseNet201 | **0.933** | 0.932 | 0.967 | 0.937 | 0.95 | 0.935 |
| | SqueezeNet | 0.861 | 0.859 | 0.93 | 0.87 | 0.895 | 0.865 |
| Bacterial and Viral Pneumonia | AlexNet | 0.90 | 0.94 | 0.845 | 0.86 | 0.89 | 0.921 |
| | ResNet18 | 0.87 | 0.92 | 0.82 | 0.83 | 0.87 | 0.873 |
| | DenseNet201 | **0.95** | 0.96 | 0.94 | 0.95 | 0.952 | 0.952 |
| | SqueezeNet | 0.83 | 0.905 | 0.75 | 0.79 | 0.83 | 0.84 |

The authors have also compared the work in the paper with the results of recently published works in the same problem. Vikash et al. [21] reported the detection of pneumonia using deep learning technique called CXNet-m1, which showed sensitivity, precision and accuracy of 99.6%, 93.28%, and 96.39%, respectively. Krishnan et al. [29] evaluated the performance of VGG16 CNNs to classify pneumonia into normal, bacterial and viral pneumonias. This showed sensitivity, precision and accuracy of 99.5%, 97.7%, and 96.2%, respectively for pneumonia detection; however, 98.4%, 92% and 93.6%, respectively for distinguishing viral and bacterial pneumonias. Table 5 summarized the comparison of others works in pneumonia and types of pneumonia detection from chest X-ray images.

**Table 5:** Comparison with other recent similar works.

| Author | Classes | Technique | Image no. | Recall /Sensitivity (%) | Precision (%) | AUC (%) | Accuracy (%) |
|---|---|---|---|---|---|---|---|
| Vikash et al.[21] | Normal & Pneumonia | Different pre-trained CNN model | 5232 | 99.6 | 93.28 | 99.34 | 96.39 |
| Sivarama krishnan et al. [29] | Normal & Pneumonia | customized VGG16 CNN model | 5856 | 99.5 | 97.0 | 99.0 | 96.2 |
| Kermanye tal.[54] | Normal & Pneumonia | Inception V3 pretrained CNN model | 5232 | 93.2 | 90.1 | - | 92.8 |
| [1]*M.Togacar et al. [22] | Normal & Pneumonia | Deep CNN model | 5849 | 96.83 | 96.88 | 96.80 | 96.84 |
| A. A. Saraiva et al. [23] | Normal & Pneumonia | Neural network | 5840 | 94.5 | 94.3 | 94.5 | 94.4 |
| Enes AYAN et al [24] | Normal & Pneumonia | VGG16 deep learning model | 5856 | 89.1 | 91.3 | 87.0 | 84.5 |
| Xianghong Gu et al.[30] | Bacterial & Viral Pneumonia | Deep Convolutional neural network | 4882 | 77.55 | 88.86 | 82.3 | 80.4 |
| [1]*Sivarama krishnan et al. [29] | Bacterial & Viral Pneumonia | customized VGG16 CNN model | 3883 | 98.4 | 92.0 | 96.2 | 93.6 |
| Archit Khatri et al.[25] | Bacterial & Viral Pneumonia | EMD approach | 144 | 89.5 | 80.0 | 88.0 | 83.30 |
| [1]**This | Normal & | Different | 5247 | 99.0 | 97.0 | 98.0 | 98.0 |

[1]*The best performing algorithms reported in the literature while **shows the performance of this study.

| work | Pneumonia | pre-trained CNN model | | | | |
| --- | --- | --- | --- | --- | --- | --- |
| ¹**This work | Bacterial & Viral Pneumonia | Different pre-trained CNN model | 5247 | 96.0 | 95.0 | 95.2 | 95.0 |

It is evident that DesneNet201 exhibits the highest accuracy than all the recent works in the best of our knowledge, which could be useful in developing a prototype that can automatically classify results into normal, bacterial and viral pneumonia. Training the network using a larger database and working on an ensemble of the pre-trained CNN algorithms might increase the detection accuracy, which can be done as a future work.

## 5. Conclusion

This work presents deep CNN based transfer learning approach for automatic detection of pneumonia and its' classes. Four different popular CNN based deep learning algorithms were trained and tested for classifying normal and pneumonia patients using chest x-ray images. It was observed that DenseNet201 outperforms other three different deep CNN networks. The classification accuracy, precision and recall of normal and pneumonia images, bacterial and viral pneumonia images, and normal, bacterial and viral pneumonia were (98%, 97%, and 99%); (95%, 95% and 96%) and (93.3%, 93.7% and 93.2%) respectively. There are millions of children who die each year due to this potentially fatal disease. Timely intervention with proper treatment plan through correct diagnosis of the disease can save a significant number of lives. Due to the huge number of patients attending out-door or emergency specifically in the third world countries, medical doctor's time is limited and computer-aided-diagnosis (CAD) can save lives. Moreover, there is a large degree of variability in the input images from the X-ray machines due to the variations of expertise of the radiologist. DenseNet201 exhibits an excellent performance in classifying pneumonia by effectively training itself from a comparatively lower collection of complex data such as images, with reduced bias and higher generalization. We believe that this computer aided diagnostic tool can significantly help the radiologist to take clinically more useful images and to identify pneumonia with its type immediately after acquisition. This fast classification will open up other avenues of application for this CAD tool, more particularly in the airport screening of pneumonia patients.

**Author Contributions:** Experiments were designed by TR and MEHC. Experiments were performed by TR, KFI and KRI. Results were analyzed by TR, MEHC, ZBM and MAK. All authors were involved in interpretation of data and paper writing.
**Funding:** The publication of this article was funded by the Qatar National Library and Qatar National Research Fund (QNRF) with the grant (NPRP12S-0227-190164).
**Acknowledgments:** The authors would like to thank the Qatar National Research Fund (QNRF) for the grant (NPRP12S-0227-190164) to bear the research personnel cost, which made this work possible.
**Conflicts of Interest:** The authors declare no conflict of interest.